\documentstyle[aps,prl,preprint,epsf,tighten]{revtex}

\newcommand{\beq}{\begin{equation}}
\newcommand{\eeq}{\end{equation}}
\newcommand{\beqa}{\begin{eqnarray}}
\newcommand{\eeqa}{\end{eqnarray}}

\begin{document}

\hfill CRN 95-38

\hfill TK 95 30

\hfill hep-ph/9512nnn

\bigskip\bigskip

\begin{center}

{\large\bf CHIRAL SYMMETRY AND THE REACTION \boldmath{$\gamma p\to
    \pi^0 p$}}

\end{center}

\vspace{.5in}

\begin{center}
{\large V. Bernard$^a$, N. Kaiser$^b$, Ulf-G. Mei{\ss}ner$^c$}

\bigskip

\bigskip

$^a$Laboratoire de Physique Th\'eorique,
Institut de Physique \\ 3-5 rue de l'Universit\'e, F-67084 Strasbourg
Cedex, France\\
Centre de Recherches Nucl\'eaires, Physique Th\'eorique\\
BP 28, F-67037 Strasbourg Cedex 2, France\\
email: bernard@crnhp4.in2p3.fr\\
\vspace{0.3cm}
$^b$Technische Universit\"at M\"unchen, Physik Department T39\\
James-Franck-Stra\ss e, D-85747 Garching, Germany\\
email: nkaiser@physik.tu-muenchen.de\\
\vspace{0.3cm}
$^c$Universit\"at Bonn, Institut f{\"u}r Theoretische Kernphysik\\
Nussallee 14-16, D-53115 Bonn, Germany\\
email: meissner@pythia.itkp.uni-bonn.de\\
\end{center}


\thispagestyle{empty}

\begin{abstract}
We analyze the new threshold data for neutral pion photoproduction off protons
in the framework of heavy baryon chiral perturbation theory.
We show that large loop corrections are needed to understand the
S--wave multipole $E_{0+}$ and that all pertinent low--energy constants
can be understood within the framework of resonance exchange
saturation. Previous inconsistencies in the description of this
reaction in the threshold region are resolved.
\end{abstract}

\vfill

\today

\newpage

\noindent Neutral pion photoproduction off  protons has been a hot
topic ever since the Saclay \cite{mazz} and Mainz \cite{beck} groups
claimed a sizeable deviation from a so--called low--energy theorem
(LET) for the electric dipole amplitude $E_{0+}$
derived in 1970 \cite{vz} \cite{deBaenst}. However,
reexaminations of these data seemed to bring the empirical value in
agreement with the theoretical prediction, see
e.g. \cite{bh} \cite{berg}. On the theoretical side, it was shown that
the low--energy theorem of Refs.\cite{vz} \cite{deBaenst} is indeed
incomplete \cite{bgkm} and that the expansion of the electric dipole
amplitude in powers of $\mu = M_\pi /m$ (with $M_\pi$ and $m$ the pion
and the nucleon mass, respectively) is slowly converging and therefore
hard to pin down accurately. The numerical closeness of the empirical
value of $E_{0+}$ at threshold with the one based on the
incomplete LET has led to
a flurry of proposals to reinterpret or resurrect the latter (for a
detailed discussion, see e.g. \cite{cnpp}). Two new developments,
however, allow us to show in this letter that indeed there is no
mystery about the threshold data for $\gamma p \to \pi^0 p$ if one
performs a sufficiently accurate calculation in chiral perturbation
theory. First, the theoretical framework to do just that
was laid out in Ref.\cite{bkmz}, as
discussed very briefly below. Second, the new data from the
TAPS collaboration have now been released \cite{fuchs} and they
show some discrepancies to the previously considered best data of Beck
et al.\cite{beck} \cite{remark1}.

Let us briefly review the pertinent results of Ref.\cite{bkmz}. In
that paper, heavy baryon chiral perturbation theory \cite{jm} was used
to calculate the S--wave multipole $E_{0+}$ to order $q^4$ (where $q$
denotes a small momentum) and the P--wave multipoles $P_{1,2,3}$
to order $q^3$ \cite{remark2}. The pertinent effective Lagrangian
takes the form
\beq
{\cal L}_{\pi N} = {\cal L}_{\pi N}^{(1)} + {\cal L}_{\pi N}^{(2)} +
{\cal L}_{\pi N}^{(3)} + {\cal L}_{\pi N}^{(4)}+ {\cal L}_{\pi \pi}^{(2)}
\label{leff}
\eeq
where the superscript $(i)$ refers to the number of derivatives or
meson mass insertions. The structure of ${\cal L}_{\pi N}$ is
discussed in detail in the review \cite{bkmr} and a pedagogical
introduction can be found in \cite{prag}.
Besides the loop and pseudovector (pv) Born contributions,
there are to this order
two counter terms in the S--wave and one in the P--wave $P_3$,
\beq
E_{0+}(\omega) = E_{0+}^{\rm Born}(\omega) + E_{0+}^{\rm loop}(\omega)
+ e \, a_1 \, \omega \, M_\pi^2 + e \, a_2 \, \omega^3 \, \, ,
\label{e0p}
\eeq
\beq
P_i (\omega) = P_i^{\rm Born}(\omega) + P_i^{\rm loop}(\omega) \quad i
= 1,2 \, \, ,
\label{p12}
\eeq
\beq
P_3 (\omega) = P_3^{\rm Born}(\omega) + e \, b_P \, \omega |\vec{q} \, | \, \,
\label{p3}
\eeq
with $\omega$ the pion energy in the cms system, $\vec{q}$ the pion
momentum and $e^2/4\pi = 1/137.036$. We have
not made explicit
the scale dependence of the low--energy constants $a_1$ and $a_2$.
In what follows, we use $\lambda = m$.
At threshold, we have $\omega_0 = M_{\pi^0} = 134.97$ MeV.
The pv Born contributions include
the coupling proportional to the anomalous magnetic moment of the
proton, $\kappa_p$. In the chiral counting, these stem from the
dimension two Lagrangian ${\cal L}_{\pi N}^{(2)}$. Based on the data of
Ref.\cite{beck}, the three low--energy constants $a_{1,2}$
 and $b_P$  could be determined by a best fit.
First, the numerical value for $b_P$ can be estimated from resonance
exchange \cite{reso}, in this case from the $\Delta$ (in the static
isobar model) and the
vector mesons, $V= \rho^0 + \omega$,
\beq
b_P^{\rm reso} = b_P^\Delta + b_P^V = (9.7 + 3.1)\,{\rm GeV}^{-3} \, \,
\label{bpr}
\eeq
which shows that the vector meson contribution can not be neglected
\cite{remark3}. The fitted value for $b_p$ is close to the number
given in Eq.(\ref{bpr}). However, letting the two S--wave constants
$a_1$ and $a_2$ completely free, they turn out to be very large in
magnitude but of different sign. If one restricts these coefficients
again by resonance exchange, one can only vary the $\Delta$ off-shell
parameters within some bounds \cite{benm}  and finds much smaller
values for $a_1$ and $a_2$ (typically a factor 20 smaller than in the
free fit). This signals that there are either enormous higher order
corrections or that the strong energy dependence of $E_{0+}(\omega)$
as suggested by the data of \cite{beck} is incorrect.
It is important to note, however, that the sum $a_1 + a_2$ is
roughly the same in both procedures. In effect, if no anomalously
large coefficients appear, only this sum plays a role (in the
threshold region). We also remark
that the form of Eqs.(\ref{p12}) has lead to novel P--wave LETs
for $P_1$ and $P_2$. These
will be tested directly in polarization measurements at MAMI soon
(for a somewhat
model--dependent analysis, see \cite{berg2}).

We can now use this
formalism to analyze the {\it new} TAPS  data of Fuchs et al. \cite{fuchs}.
In Fig.1, we show the fit constrained by resonance exchange for the
differential cross sections and in Fig.2 for the total cross section.
For the differential cross sections, all data up to $E_\gamma =160$
MeV were used in the fit but only the ones up to $E_\gamma = 152$ MeV
are shown in Fig.1.
Before discussing our fit parameters, we note that the
differential cross sections above the $\pi^+ n$ threshold show much
less of the pronounced bell shape as inferred from the older data and
that the total cross section is somewhat decreased, which is of
particular relevance for the extraction of $E_{0+}$. Let us now
discuss in more detail the fits based on the theoretical framework
of Ref.\cite{bkmz}.
For the $\Delta$, we keep the two $\gamma N \Delta$ couplings fixed, $g_1
= g_2 = 5$. The off---shell parameter $Y$ is severely constrained by
the $\Delta$ contribution to the magnetic polarizability of the
proton, we set $0.1 \le Y \le 0.14$ , i.e. $6.4 \le \delta \beta_p^\Delta
\le 7.5$ (in units of 10$^{-4}$~fm$^3$) \cite{mnz} \cite{bkms}. Also,
$Z$ is bounded by the $\Delta$ contribution to the $\pi N$ scattering
volume $a_{33}$, $-0.4 \le Z \le -0.2$ \cite{armin}. $X$ is varied
within the range given in \cite{benm}. We find $X=2.75$
\cite{remarkx}, $Y=0.10$ and
$Z = -0.21$ which translates into
\beq
(a_1+a_2)^{V + \Delta} = (2.67 + 3.92)  \, {\rm GeV}^{-4} =
 6.59 \, {\rm GeV}^{-4} \, , \, \, \, \,
b_P^{\rm reso} = 13.0 \, {\rm GeV}^{-3} \, .
\label{a12bpr}
\eeq
The $\chi^2$/dof is 2.21.
We remark that the value for $b_P$ does indeed nicely agree
with the resonance saturation estimate, Eq.(\ref{bpr}). Note
that $b_P^{\rm reso}$ does not depend on $X$ and its possible values
are strongly constrained by the ranges of $Y$ and $Z$ discussed above.
It is thus gratifying that one can obtain such a consistent
description of this low--energy constant.
The sum $(a_1 + a_2)$ is consistent with the free fit value
of 6.60 GeV$^{-4}$ (we do not show the free fit since it is
essentially the same as the resonance one). We also note that
the resonance fit of Ref.\cite{bkmz} already had
$a_1 + a_2 = 6.67$ GeV$^{-4}$. The apparent mismatch between the
free and the resonance fit discussed in \cite{bkmz} has turned out to
be an artefact related to the old data.

In Fig.~3, we show the electric dipole amplitude $E_{0+}$. Its
 values at the $\pi^0 p$ and the $\pi^+ n$ threshold are
\beq
E_{0+} (\pi^0 p) = -1.16 \cdot 10^{-3}/M_{\pi^+} \, , \quad
E_{0+} (\pi^+ n) = -0.44 \cdot 10^{-3}/M_{\pi^+} \, \, ,
\label{e0pval}
\eeq
to be compared with
$E_{0+}^{\rm exp} (\pi^0 p) = -1.31 \pm 0.08 \cdot 10^{-3}/M_{\pi^+}$
\cite{fuchs} and $E_{0+}^{\rm exp} (\pi^+ n) \simeq -0.4
\cdot 10^{-3}/M_{\pi^+}$ (as read off from Fig.4 of \cite{fuchs}).
The value of $a_1+a_2$ in Eq.(\ref{a12bpr}) amounts to an
$E_{0+}$-contribution of $+0.3$ from vector mesons and $+0.4$ from the
$\Delta$ (in units of $10^{-3}/M_{\pi^+}$).  Almost the same number
for the sum of vector meson and nucleon
resonance countributions to $E_{0+}$ at threshold is
reported in Ref.\cite{dt} (see also \cite{david}).
In the threshold region, the shape
for $E_{0+} (\omega)$ shown in Fig.~3 can be well represented by a
two--parameter fit of the form
(as discussed in some detail in Ref.\cite{bkmz})
\beq
E_{0+} (\omega) = -a - b \, \sqrt{1 - \omega^2 / \omega_c^2 } \, \, ,
\label{e0pform}
\eeq
with $\omega_c = 140.11$ MeV the pion energy at the $\pi^+ n$ threshold.
We find $a = 0.44 \cdot 10^{-3} / M_{\pi^+}$, $b = 2.9 \cdot
 10^{-3}/M_{\pi^+}$
and $b_P = 12.9$ GeV$^{-3}$. The $\chi^2$/dof is 2.22, i.e. almost
identical to the one of the resonance fit. This indicates that the
mild slope of Re~$E_{0+} (\omega )$ behind the $\pi^+ n$ threshold is
not significant.  The value for $b$ is somewhat below the
one estimated from the Fermi--Watson theorem, $b^{\rm FW} =
3.7 \cdot 10^{-3}/M_{\pi+}$. Note, however, that this  is based on the
assumption of exact isospin symmetry, whereas the clearly visible cusp
effect in $E_{0+} (\omega)$ is due to the pion mass difference,
i.e. an isospin--violating effect. A more consistent treatment of such
effects is certainly needed. For a study of the cusp effect in
$E_{0+}$ in terms of a multi--channel S--matrix, see Ref.\cite{aron}.
 We note that the small value of $a$ is a
clear indication of chiral loops - with simple pseudovector Born terms
one can not get such a small value (this was already
stressed in \cite{bkmz}). $\gamma p \to \pi^0 p$ has also been
remeasured at Saskatoon \cite{jackpr}.
Between $\pi^0 p$ and $\pi^+ n$ thresholds,
the new SAL data are consistent with the ones of Ref.\cite{fuchs}. For
larger energies, however, the new SAL data agree with the older Mainz
data \cite{beck}. This does not affect the threshold value of $E_{0+}$
but rather leads to a larger value of $b_P$. The experimental
discrepancy remains to be clarified.

For the respective slopes of the P--wave multipoles,
the LETs together with the best
resonance fit give $P_1 / | {\vec q} \,| = 0.480$~GeV$^{-2}$,
$P_2 / | {\vec q} \,| = -0.512$~GeV$^{-2}$
and $P_3 / | {\vec q} \,| = 0.544$~GeV$^{-2}$,
i.e. all P--waves $P_{1,2,3}$ are of the
same magnitude close to threshold. Consequently, the photon asymmetry
$\Sigma(\theta )$ to be measured at MAMI is expected to be small in
the threshold region since $\Sigma \sim (|P_3|^2 - |P_2|^2)$.

To summarize, we have shown that within the framework of
chiral perturbation theory, one is able to consistently understand
the new threshold data for the reaction $\gamma p
\to\pi^0 p$. Loop effects are clearly visible in the
S--wave. Furthermore, the three low--energy constants are fully
understood within the framework of resonance saturation. In contrast
to common folklore, these resonances pose no problem and do not have
to be treated as dynamical degrees of freedom (as long as one stays in
the threshold region). In the next step, this formalism should be
extended to electroproduction to discuss the new data from NIKHEF
\cite{benno} and MAMI \cite{distler}.

\vspace{1.5cm}

\section*{Acknowledgements}

We are grateful to Michael Fuchs for providing us with the TAPS data
before publication and to Jack Bergstrom for informing us
about the SAL results.




\newpage

$\,$\bigskip

\section*{Figures}

\begin{enumerate}

\item[Fig.1] Differential cross sections in the threshold
  region (in nb/sr) for lowest 9 values of the photon lab energy $E_\gamma$
  versus the cm scattering angle $\theta$. The solid line is the best
  resonance fit, the data are from \cite{fuchs}.

\item[Fig.2] Total cross section in the threshold region (in $\mu$b)
  versus $E_\gamma$. For  notations, see Fig.1.

\item[Fig.3] The real part of the electric dipole amplitude
  in the threshold region.

\end{enumerate}

\newpage

$\,$\vspace{2cm}

\epsfxsize=15cm
\epsfysize=18cm
\epsffile{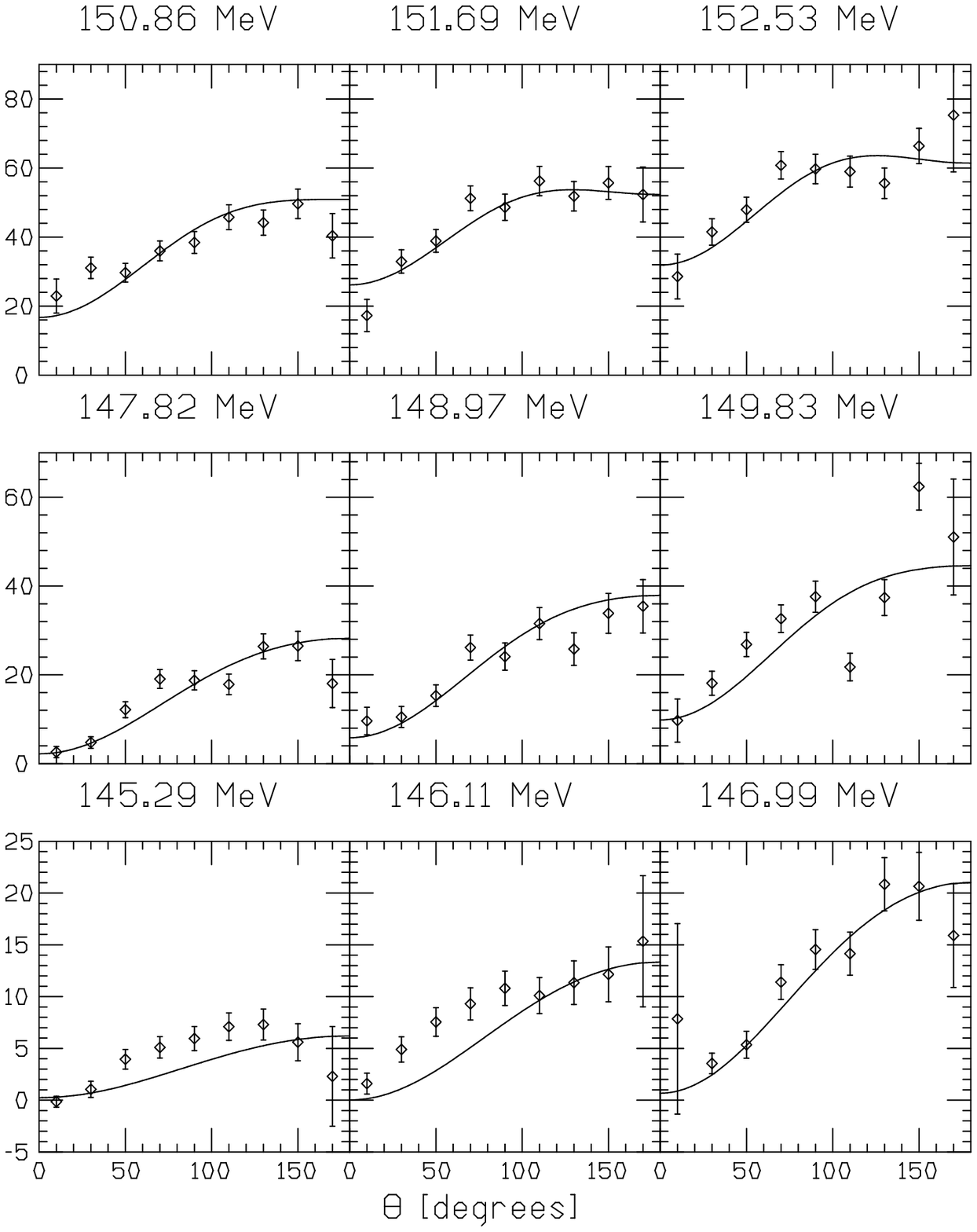}

\vspace{1cm}

\centerline{ \large{Figure 1}}

%
%
%
%

\newpage

$\,$\vspace{0.5cm}

\hskip 1.5in
\epsfxsize=6.5cm
\epsffile{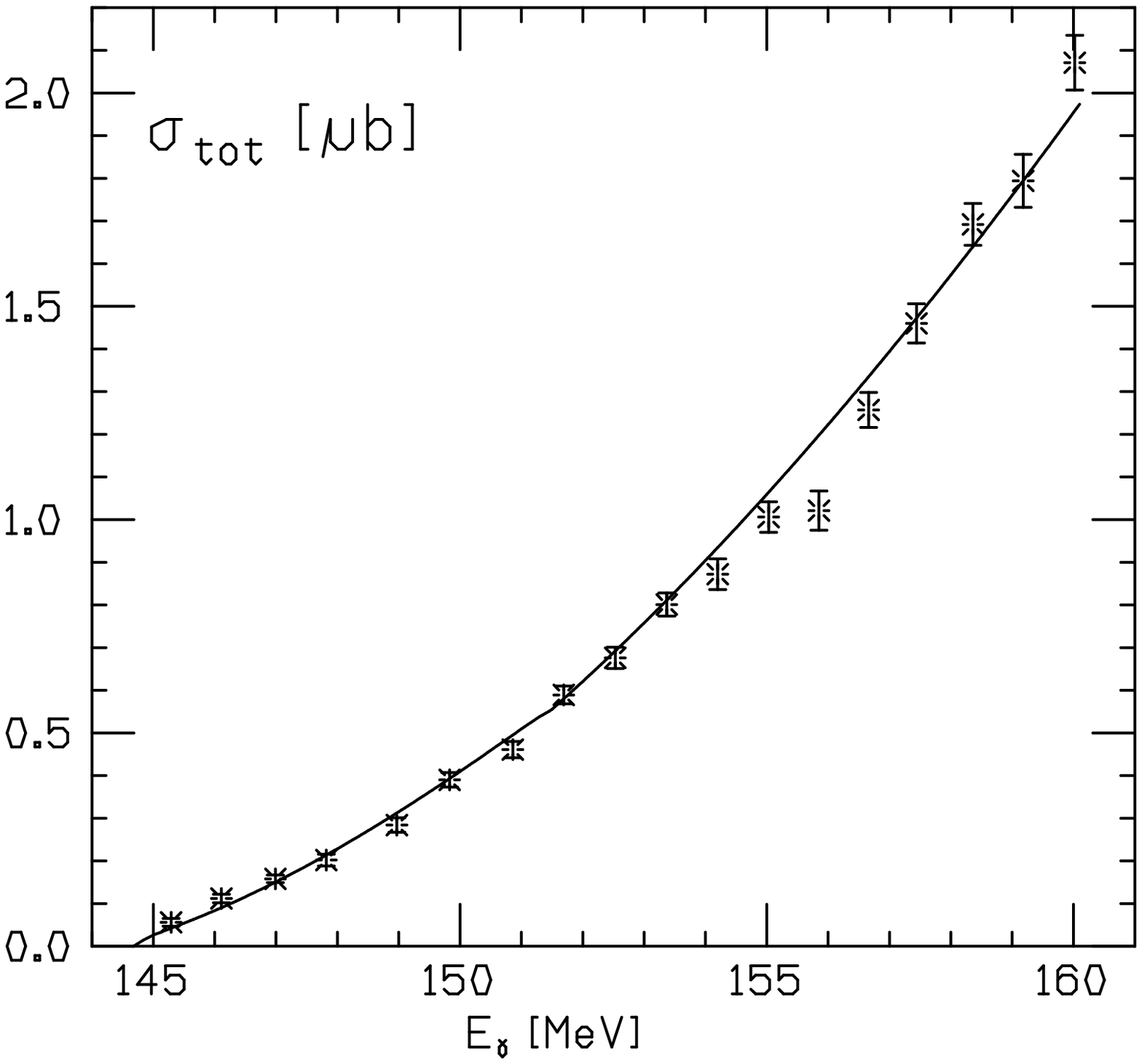}


\centerline{ \large{Figure 2}}

$\,$\vspace{1.5cm}

\hskip 1.5in
\epsfxsize=6.5cm
\epsffile{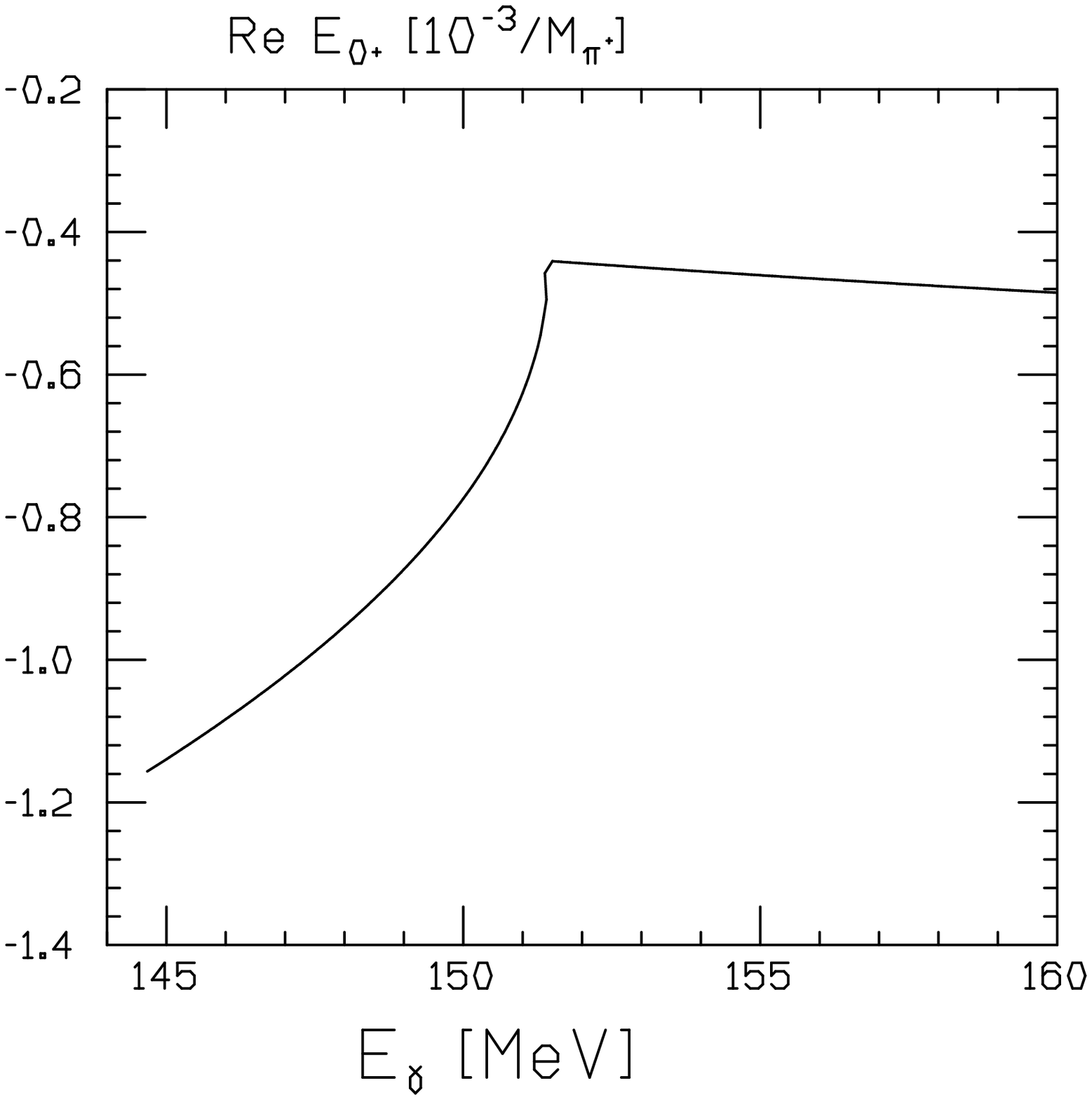}


\centerline{ \large{Figure 3}}


\begin{thebibliography}{99}
\bibitem{remark1} These discrepancies are discussed in
  Ref.\cite{fuchs}.
\bibitem{remark2} These multipoles are related to the more commonly
  used electric and magnetic ones via $P_1 = 3E_{1+} + M_{1+} -
  M_{1-}$, $P_2 = 3E_{1+} - M_{1+} + M_{1-}$ and $P_3 = 2M_{1+}
  +M_{1-}$. They appear naturally in the T--matrix for $\pi^0$ production.
\bibitem{remark3} Notice that we drop here the $M_\pi^2$ term in the
denominator of Eq.(4.10) of Ref.\cite{bkmz}. This is consistent within
the power counting and also discussed in that paper.
\end{thebibliography}
\end{document}